\documentclass{aa}  

\pdfoutput=1
\usepackage{graphicx}
\usepackage{txfonts}
\usepackage{hyperref}
\usepackage{amsmath,amstext}
\usepackage{gensymb}
\usepackage{natbib}
\bibpunct{(}{)}{;}{a}{}{,} 

\begin{document}

\title{A population of high-velocity absorption-line systems residing in the Local Group}

\author{S. J. D. Bouma \and P. Richter \and C. Fechner}
\institute{Institut f\"ur Physik und Astronomie, Universit\"at Potsdam, Karl-Liebknecht-Str. 24/25, 14476 Golm, Germany\\
\email{sbouma@astro.physik.uni-potsdam.de}}

\abstract
{}
{We aim to investigate the ionisation conditions and distances of Galactic high-velocity clouds (HVCs) in the Galactic halo and beyond in the direction of the Local Group (LG) barycentre and anti-barycentre, by studying spectral data of 29 extragalactic background sources obtained with {\it Cosmic Origins Spectropgraph} (COS) installed on the Hubble Space Telescope (HST).}
{We model column-densities of low, intermediate, and high ions, such as Si\,{\sc ii}, C\,{\sc ii}, Si\,{\sc ii}, Si\,{\sc iv}, and C\,{\sc iv} and use this to construct a set of Cloudy ionisation models.}
{In total, we found 69 high-velocity absorption components along the 29 lines of sight. The ones in the direction of the LG barycentre span the entire range of studied velocities, $100 \lesssim |v_{LSR}|\lesssim 400$ km\,s$^{-1}$, while the anti-barycentre sample has velocities up to about 300 km\,s$^{-1}$. For 49 components, we infer the gas densities. In the direction of the LG barycentre, the gas densities exhibit a large range between log\,$n_{\rm H}=-3.96$ to $-2.55$, while in the anti-barycentre direction the densities are systematically higher, log\,$n_{\rm H}>-3.25$. The barycentre absorbers can be split into two groups based on their density: a high density group with log\,$n_{\rm H}>-3.54$, which can be affected by the Milky Way radiation field, and a low density group (log\,$n_{\rm H} \leq -3.54$). The latter has very low thermal pressures of $P/k<7.3$ K\,cm$^{-3}$.}
{Our study shows that part of the absorbers in the LG barycentre direction trace gas at very low gas densities and thermal pressures. Such properties indicate that these absorbers are located beyond the virial radius of the Milky Way. Our study also confirms results from earlier, single-sightline studies, suggesting the presence of a metal-enriched intragroup medium filling the LG near its barycentre.
}


\maketitle

\section{Introduction}

Since the first discovery of the so-called high-velocity clouds (HVCs) by \citet{Muller63} in the Milky Way (MW) halo, their origin has been a matter of debate. These HVCs are seen in 21cm emission or in ultraviolet or optical absorption at radial velocities |v$_{LSR}$| $>90-100$ km\,s$^{-1}$. Already more than 50 years ago, \citet{Oort66} gave a list of possible origins, such as them being parts of supernova shells, intergalactic gas that is being accreted by the Milky Way, small satellite galaxies of the Milky Way or independent galaxies of the Local Group (LG). As we know today, most of the HVCs have low metallicities ($<0.3$ solar, typically; e.g. \citet{Gibson00,Wakker01,Richter01,Collins07,Fox10,Shull11}), indicating that they indeed represent gas accreted from outside, either from satellite galaxies or the intergalactic medium (IGM). The Magellanic Stream (MS), for example, is the result of the gravitational and hydrodynamical interaction between the Milky Way and the Magellanic Clouds \citep[e.g.][]{Putman98}. 

Halo clouds with somewhat lower velocities, namely the intermediate-velocity clouds (IVCs; $40-50<$ |v$_{\rm LSR}|<90-100$~km\,s$^{-1}$), often appear to be associated with a so-called Galactic fountain. In the Galactic fountain model \citep{Shapiro76,Fraternali08}, gas is blown out by superbubbles (created by SNe explosions), mixes with the ambient coronal gas, cools, and falls back onto the disk. The relatively high metallicity found in many IVCs \citep[and references therein]{Wakker01} further supports a Galactic origin of the IVCs as part of the Galactic fountain model. This model, however, does not explain HVCs, thus making it likely that their origin is not within the disc or bulge of the MW \citep{Marasco12}. Some HVCs may be explained by galactic rotation, but only when systematic peculiar velocities are assumed. An example is the Outer Arm HVC being an extension of the Outer Arm \citep{Haud92}.

One of the keys to our understanding of the origin and nature of HVCs is the knowledge of their distances from us. Most HVCs are located within 20~kpc and relatively close to the Milky Way disk \citep{Wakker01, Thom06, Thom08, Wakker07, Wakker08}, but some clouds possibly lie outside of the virial radius of the MW \citep{Sembach03, Richter17}. These faraway clouds could be evidence that some HVCs have a Local Group origin. \citet{Blitz99} proposed that some of the compact HVCs (CHVCs) represent distant (extragalactic) clouds confined by dark matter mini-halos that are moving towards the LG barycentre. This theory has been disputed, however, as there is  no stellar component found in those HVCs \citep{Simon02,Hopp03} and the nearby M31 does not show any significant CHVC population at large galactocentric radii \citep{Westmeier08}.

In addition to distance, the kinematics of HVCs can also help in understanding their origin. As pointed out by \citet{Nicastro03}, high-velocity O\,{\sc vi} absorption in the MW halo has negative velocities for absorbers located between $0\degree\leq l \leq 180\degree$, while the systems on the other half of the hemisphere have mostly positive velocities. This trend disappears when velocities are transformed to the Local Group Standard of Rest (LGSR) frame. Furthermore, using the LGSR or Galactic Standard of Rest (GSR) decreases the amplitude of the average velocity vector of the halo O\,{\sc vi} absorption when compared to the LSR. Another, much more detailed study of O\,{\sc vi} absorption in the halo \citep{Sembach03} confirms that the dispersion around the mean velocity of their absorption systematically decreases if the LSGR or GSR is taken as reference frame. These results indicate that at least a sub-set of the high-velocity O\,{\sc vi} absorbers may not be located in the MW, but rather are spatially and kinematically associated with gas filling the Local Group (LG).

Following up on this, \citet[hereafter referred to as R17]{Richter17} compared in their all-sky HVC absorption survey the absorption velocities of HVCs in the direction of the LG barycentre ($l<180\degree$, $b < -30\degree$) to those of HVCs at $l>240\degree$, $b > 60\degree$ in low and intermediate ions such as Si\,{\sc ii} and Si\,{\sc iii}. A velocity dipole is clearly visible, in which the gas in the direction of the barycentre has negative velocities up to v$_{\rm LSR}= -400$ km\,s$^{-1}$, while in the opposite direction, the velocities are positive up to  v$_{\rm LSR}=+300$ km\,s$^{-1}$. Since such high radial velocities cannot be explained by Galactic rotation, the observed kinematics of the HVC absorber population could be the result of the MW moving towards the LG barycentre and away from the anti-barycentre direction. In this scenario, gas at rest in the LG barycentre or coming from M31 would have negative velocities, as it is coming towards the MW, while LG gas in the anti-barycentre direction would be left behind, leading to positive velocities.

Nearby HVCs are not only expected to have on average lower velocities than those far away, they are also expected to have substantially higher thermal gas pressures, as they are located deep within the MW potential well, where they are pressure-confined by the ambient hot coronal gas in the outer disk-halo interface. In turn, the very low thermal gas pressures measured in earlier HVC studies in the high-velocity barycentre absorbers towards PKS\,2155$-$304 and Mrk\,509 \citep[$P/k~<~5~$K\,cm$^{-3}$;][]{Sembach99,Collins04} indicate that these clouds cannot reside in the inner halo (as they would collapse rapidly), but are located in the outer Galactic halo or even beyond the MW virial radius in the LG, where the thermal pressures are at least two orders of magnitude lower than in the inner halo \citep{Miller15,Wolfire95}.  

Finally, an LG origin of barycentre/anti-barycentre high-velocity absorbers is also supported by the ionisation structure of the absorbing gas (as indicated by the column-density ratios of low and high ions; see R17) and, in case of the barycentre clouds, the lack of a two-phase structure of the high-velocity clumps seen in H\,{\sc i} 21cm emission \citep{Winkel11}.

In this study, we investigate in more detail the physical conditions of the barycentre/anti-barycentre high-velocity absorbers identified in R17 to further explore a possible LG origin of these clouds.
For this, we re-analyse high-velocity absorption along 29 sightlines and model the absorber structure and the ionisation conditions in the gas. Our paper is organised as follows. In Sect.\,2, we describe the absorber sample and the analysis methods. In Sect.\,3, we present the results of our absorber-modelling analysis. The ionisation modelling of the absorbers is presented in Sect.\,4. In Sect.\,5, we discuss the implications of our results with regard to distance and origin of the absorbers. A summary of our study is given in Sect.\,7.


\section{Data acquisition and spectral analysis}

\begin{figure*}[htp]
\centering
\includegraphics[width=1.0\textwidth]{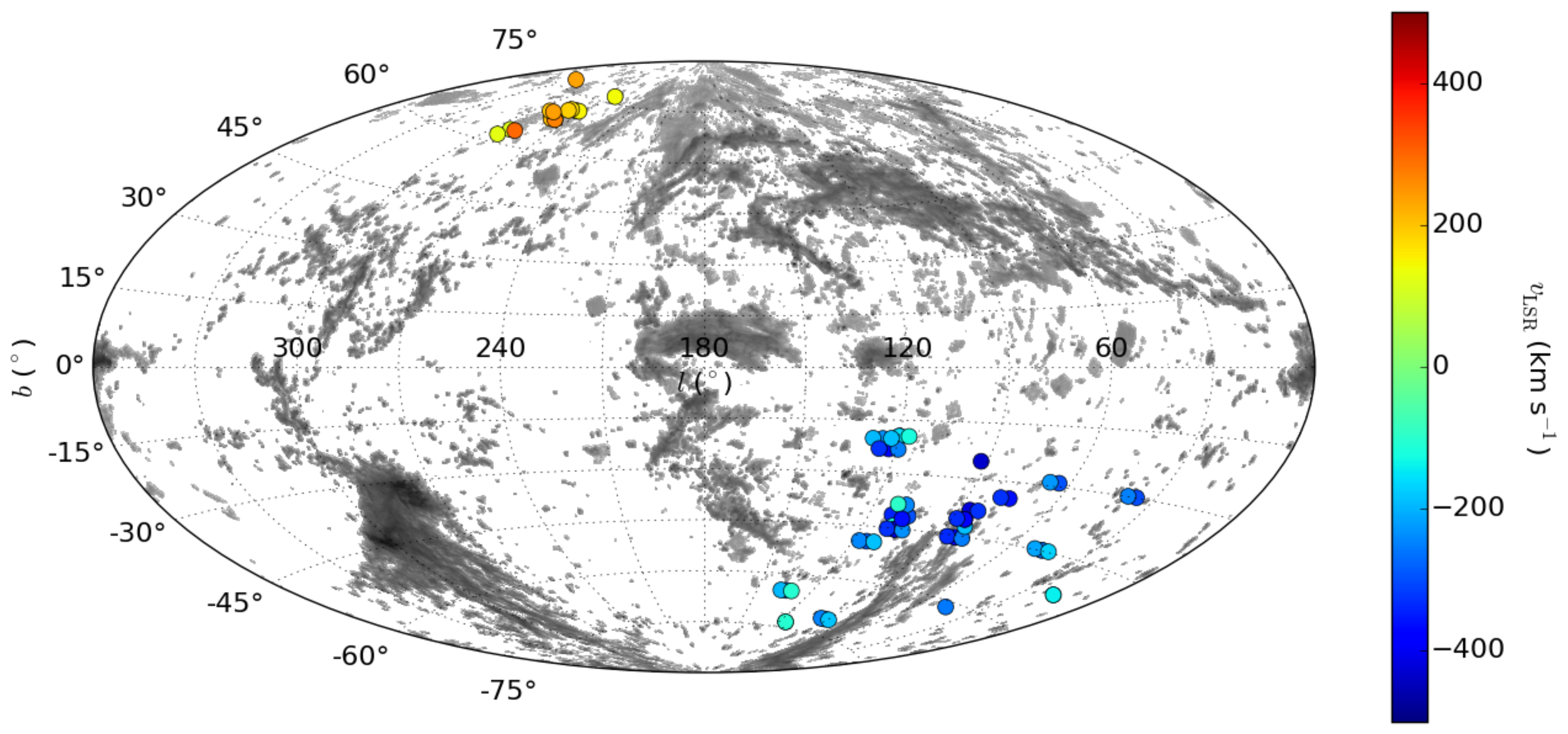}
\caption{
Sky distribution of the QSO si\textit{}ghtlines considered in this study in Galactic coordinates. The colour indicates the observed LSR velocity of the high-velocity components seen in these directions. As can be clearly seen, the high-velocity absorbers form a velocity dipole on the sky along the LG major axis (see R17). H\,{\sc i} 21\,cm data from \citet{Westmeier18} is added in greyscale.
}
\label{skydist}
\end{figure*}

The 29~AGN sightlines that are considered in this paper represent a subsample of the 262~sightlines used in the HST/COS Legacy Survey of HVCs presented in R17. For the LG barycentre direction ($l<180\degree$, $b<-30\degree$), our sample includes 19 sightlines, for the LG anti-barycentre direction ($l>240\degree$, $b>60\degree$) we have 10 sightlines available. Names and Galactic coordinates of the 19 AGN are given in
Table~\ref{tab:abundances} in the Appendix. Their sky distribution, with offsets when one sightline has several absorption components, is shown in Fig.~\ref{skydist}.

The original HST/COS data were retrieved from the HST Science Archive at the Canadian Astronomy Data Centre (CADC) and reduced following the strategy described in R17. Most of the COS spectra used here have data from the G130M and G160M gratings, but for 10 sightlines only G130M data are available.
The COS G130M grating covers a wavelength range of 1150~$-$~1450~\AA. This range includes the absorption lines of Si\,{\sc ii}, Si\,{\sc iii}, Si\,{\sc iv}, and C\,{\sc ii}.
Absorption in the important C\,{\sc iv} doublet falls into the range of the COS G160M grating, which covers $\lambda$~=~1405~$-$~1775 \AA. The resolving power for both gratings is $R=16,000-21,000$ \citep{Green2012,Fischer18}. Table~\ref{table:atomicdata} lists wavelengths and oscillator strengths of the transitions used in this study to measure the above-mentioned ions of Si and C. 
Note that we here include Si\,{\sc iv} in our analysis, which is not covered in our progenitor study (R17).


\begin{table}
\caption{Atomic data for absorption lines}      
\label{table:atomicdata}    
\centering                                  
\begin{tabular}{l l l }     
\hline\hline                      
Ion & $\lambda_0$ [\AA]  & $f$ \\   
\hline                              
  C\,{\sc ii}   & 1334.53 & 0.1284 \\
  C\,{\sc iv} 	& 1548.20 & 0.1897 \\
  C\,{\sc iv} 	& 1550.77 & 0.0947 \\
  Si\,{\sc ii} 	& 1190.42 & 0.2919 \\
  Si\,{\sc ii} 	& 1193.29 & 0.5824 \\
  Si\,{\sc ii} 	& 1260.42 & 1.1762 \\
  Si\,{\sc ii} 	& 1526.71 & 0.1328 \\
  Si\,{\sc iii}	& 1206.50 & 1.6273 \\
  Si\,{\sc iv} 	& 1393.76 & 0.5126 \\
  Si\,{\sc iv} 	& 1402.77 & 0.2541 \\
\hline                                           
\end{tabular}
\tablefoot{Data compiled from \citet{Morton03}}
\end{table}


All COS spectra were normalised using the custom-written {\tt span} code \citep{Richter11}. While in R17 we have used the apparent optical depth (AOD) method to measure velocity-integrated column densities for the various metal ions in the high-velocity absorbers, we here model in detail the velocity-component structure of the absorbers using the component-modelling method \citep{Richter13}. For this, a synthetic spectrum is constructed from a component model for each absorber, so that it matches the observations. The component model defines for each velocity sub-component of an HVC absorber the velocity centroid, the Doppler parameter/$b$-value, and the column density of the various ions as input parameters. This method is required for our purpose, as the physical conditions are expected to vary along the individual sub-component. Note that we are aiming at treating each individual absorption sub-component as an individual cloud, for which we model the local ionisation conditions.
Another advantage of using the modelling approach is that blended lines can be reproduced with a synthetic spectrum, providing additional constraints on the derived column densities and $b$ values. 

In this study, we concentrate on HVC absorbers that form the velocity dipole in the LG barycentre and anti-barycentre direction (R17). For the LG barycentre direction region at $l<180\degree$, $b<-30\degree$ we therefore consider only high-velocity absorbers with negative velocities. In the direction of the anti-barycentre at $l>240\degree$, $b>60\degree$, we consequently only consider HVCs with positive LSR velocities.

\begin{figure}[htp]
\centering
\includegraphics[width=0.4\textwidth]{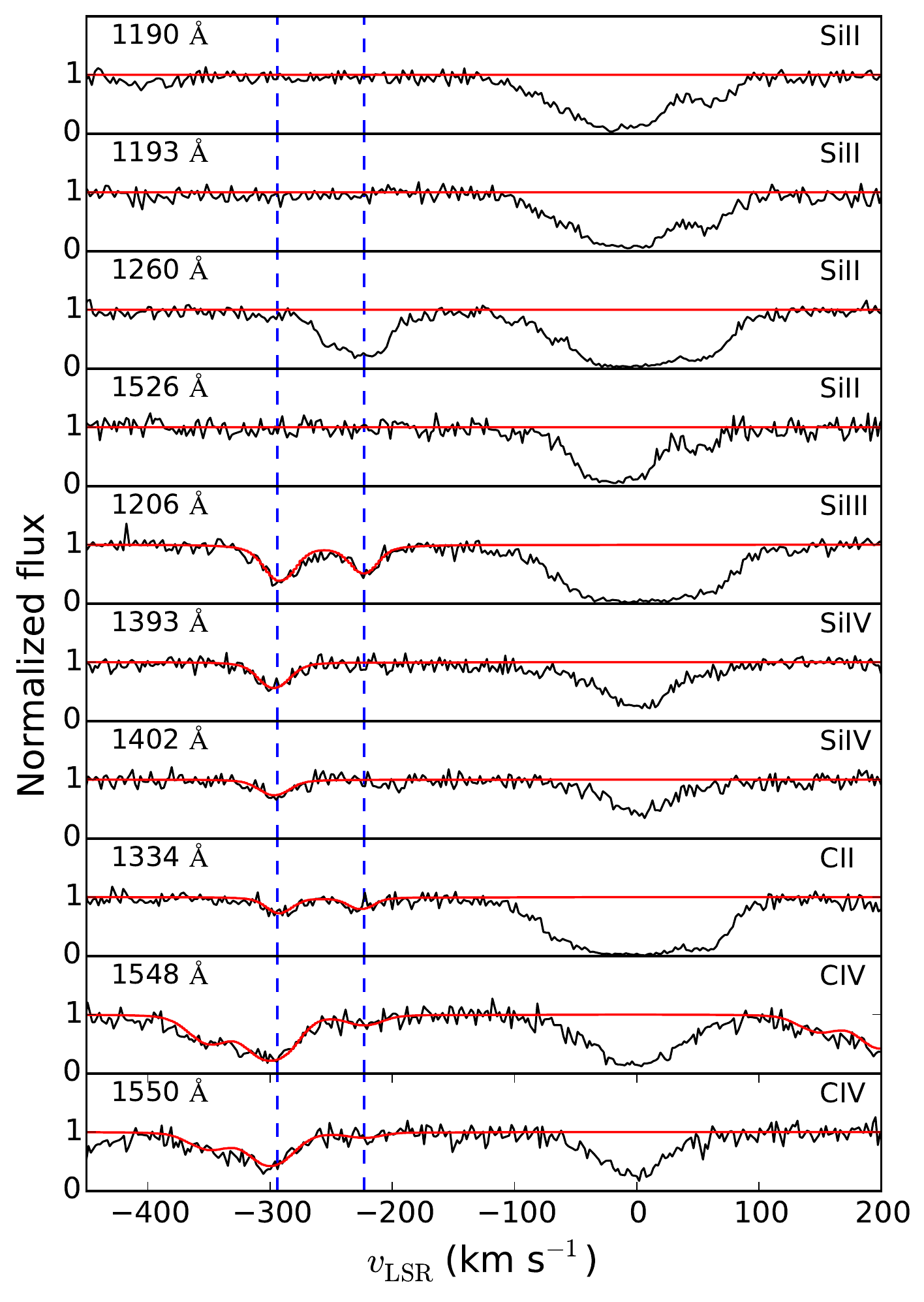}
\caption{
Example for the simultaneous modelling of high-velocity absorption components in COS spectral data. Shown is spectrum of Mrk\,1513 for the various ions considered in our study, plotted in LSR velocity space (black solid line). The synthetic model spectrum (red solid line) accurately reproduces the high-velocity absorption in the 2 high-velocity components at $-294$ and $-223$ km\,s$^{-1}$ seen in this direction, which are indicated with blue dashed lines.
}
\label{velplot}
\end{figure}

In Fig.~\ref{velplot} we show a typical example for our modelling approach for the high-velocity absorbers at $-294$ km\,s$^{-1}$ and $-223$ km\,s$^{-1}$ in the direction of the quasar Mrk\,1513. All modelling results (component velocities and column densities for each ion) are listed in Table~\ref{tab:abundances} in the Appendix. Further details regarding the modelling method can be found in \citet{Richter13}.

\section{Results from the absorption-line analysis}

Along the 19 sightlines towards the LG barycentre, 55 high-velocity absorption components with $v_{\rm LSR}\gtrsim -100$ km\,s$^{-1}$ are identified. One of these components has large uncertainties due to blending and low $S/N$ and therefore is not considered hereafter. All the sightlines in the LG barycentre direction show at least one high-velocity absorption component, with 17 sightlines having at least two absorption components. Therefore, the observed HVC sky covering fraction in this direction is 100 percent in our COS sample.

In the anti-barycentre direction, our sample of 10 sightlines exhibits 14 individual high-velocity components at $v_{\rm LSR}\geq +100$ km\,s$^{-1}$. Positive high-velocity absorption is present in all of these sightlines, implying an HVC sky covering fraction of 100 percent in our data set, with 5 of the 10 sightlines having at least 2 high-velocity absorption components.

The barycentre sightlines show more absorption components (with negative velocities) per sightline than the anti-barycentre sample does for positive velocities. 

\begin{figure}[htp]
\centering
\includegraphics[width=0.4\textwidth]{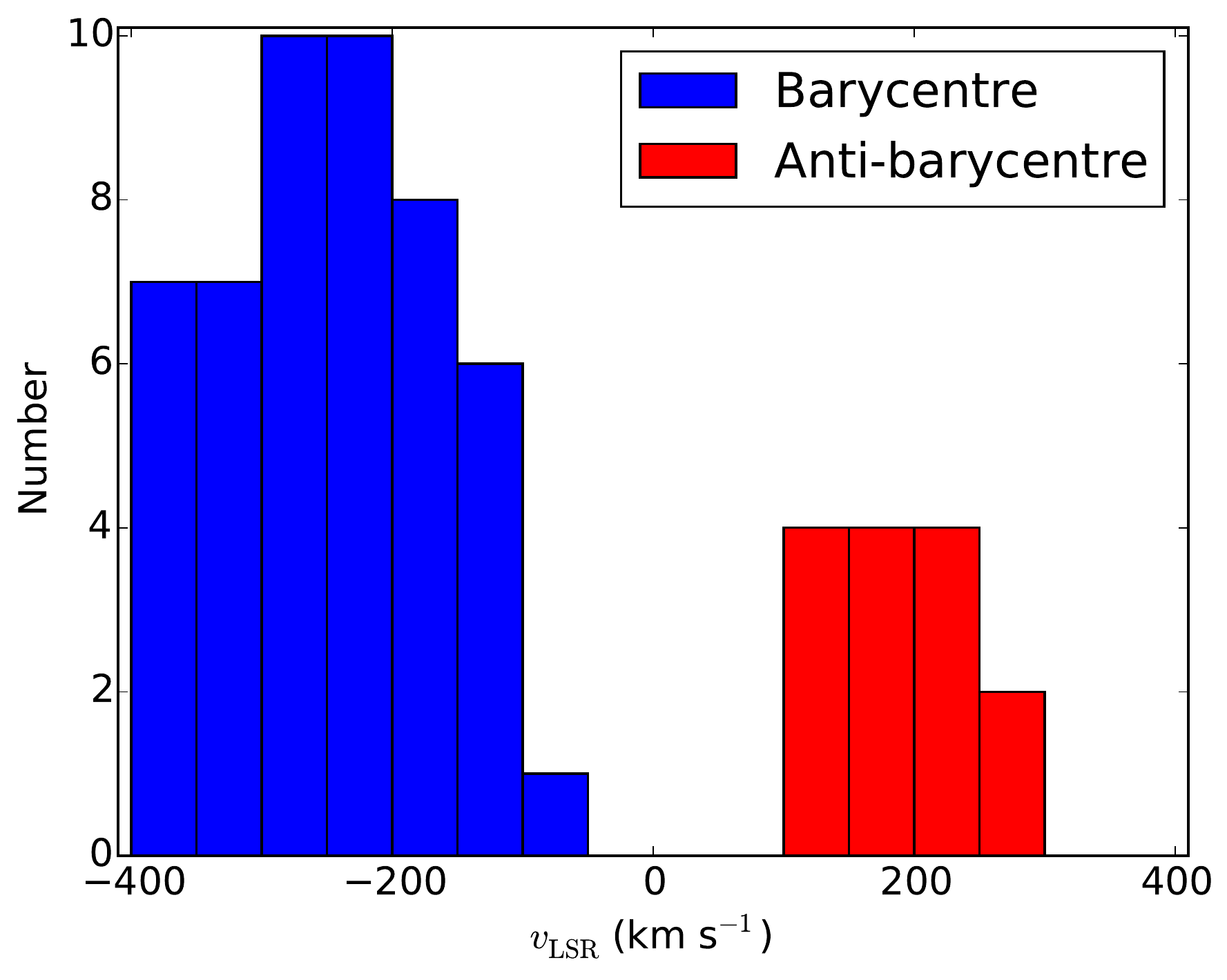}
\caption{
Velocity distribution of the detected 68 high-velocity absorption components in the LG barycentre/anti-barycentre direction. Absorbers in barycentre direction are shown in blue and those in anti-barycentre direction in red.
}
\label{vdist}
\end{figure}

In Fig.~\ref{vdist}, we show the velocity distribution of the 68 high-velocity absorption components in our sample as a histogram. Absorption in the barycentre sample has velocities spanning the entire velocity range used, from $\sim-100$ km\,s$^{-1}$ down to $-400$ km\,s$^{-1}$, while the anti-barycentre sample does not have velocities higher than about $+300$ km\,s$^{-1}$. For both samples, the absorption systems are quite evenly distributed over the velocities, although the barycentre sample has most systems in the bins between $|v_{\rm LSR}|=200-300$  km\,s$^{-1}$, which is where the anti-barycentre sample drops slightly off.


\begin{figure}[htp]
\centering
\includegraphics[width=0.4\textwidth]{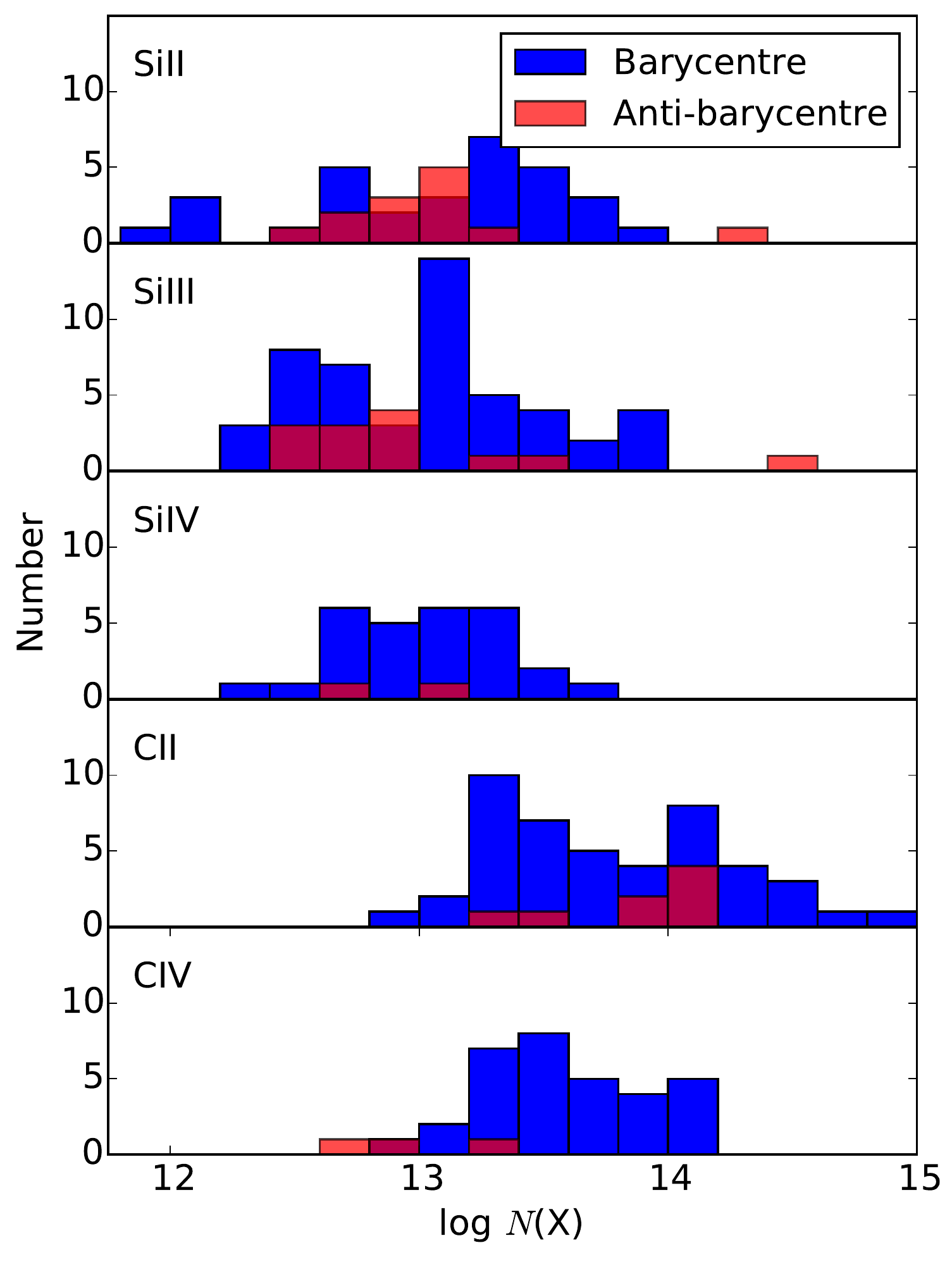}
\caption{Column-density distribution for measured ions for the LG bary-/anti-barycentre direction, with the barycentre sample in blue and the anti-barycentre sample in red.
}
\label{logNhist}
\end{figure}

In Fig.~\ref{logNhist}, we compare the column-density distribution of the barycentre-absorption components (blue) with the that of the anti-barycentre components (red) for the five different ions considered in our study. The two carbon ions that have been measured, C\,{\sc ii} and C\,{\sc iv}, span a range of slightly higher column densities (log $N$ $\sim 13-15$) than the three silicon ions Si\,{\sc ii}, Si\,{\sc iii}, and Si\,{\sc iv} (log $N$ $\sim 12-14$), owing to the higher cosmic abundance of C. Only 34 of the 51 high-velocity components for which G160M data are available show C\,{\sc iv} absorption. Likewise, only 30 of the 67 C\,{\sc ii}/Si\,{\sc ii}/Si\,{\sc iii} HVC components are detected in Si\,{\sc iv}. One HVC component is only observed in Si\,{\sc iv} and C\,{\sc iv}.
The relatively low detection rate of high ions in this sample and in the all-sky survey (R17) might be explained by the fact that they preferentially arise in low-density environments, such in the diffuse outer layers of 21cm HVCs and in diffuse halo structures far away from the Galactic plane \citep{Sembach99,Collins05}.

\begin{figure}[htp]
\centering
\includegraphics[width=0.4\textwidth]{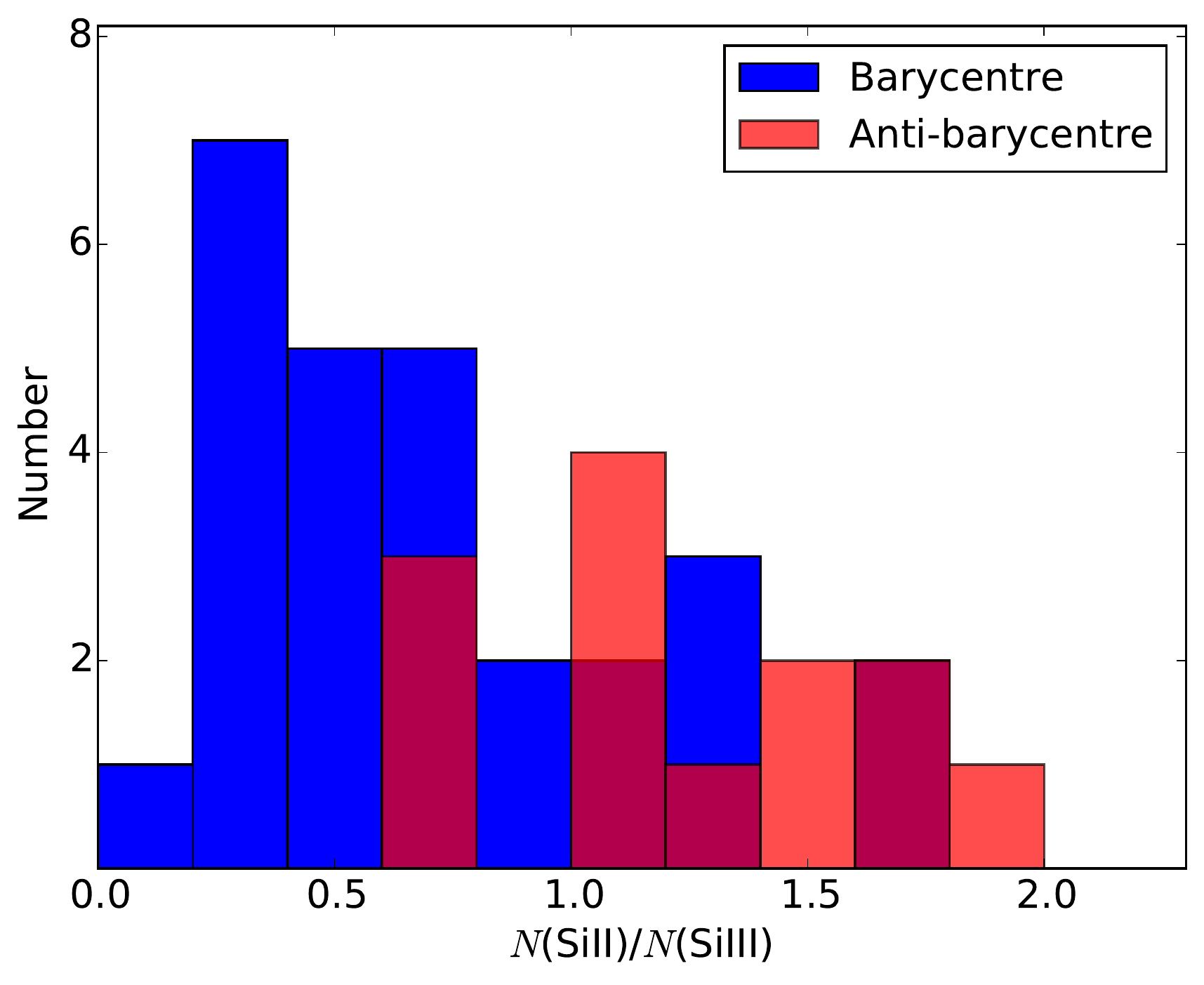}
\caption{Ratio of column densities of Si\,{\sc ii} and Si\,{\sc iii} for sightlines in the barycentre and anti-barycentre direction.
}
\label{ratioSi}
\end{figure}

The ratio Si\,{\sc ii}$/$Si\,{\sc iii} is shown in Fig.~\ref{ratioSi}. The barycentre sample shows a lower ratio than the anti-barycentre sample, indicating that the gas detected in the barycentre sample is more highly ionised. 


\section{Ionisation modelling}

To obtain more precise information about the physical conditions in the identified high-velocity absorption components we have constructed photoionisation models using the Cloudy code version 17.00 \citep{Ferland13}.

For a given ionisation model, that is usually defined by the background ionising radiation field, the absorber geometry, the neutral gas column density, the gas temperature, and the metallicity, Cloudy delivers predictions for the column densities of metal ions as a function of the gas density and ionisation parameter, $U=n_{\gamma}/n_{\rm H}$. In turn, an observed set of ion column densities in an absorber can be used together with Cloudy to constrain the local gas density. In our case, the observed ion column densities (or limits) for Si\,{\sc ii}, Si\,{\sc iii}, Si\,{\sc iv}, C\,{\sc ii}, and C\,{\sc iv} are used to constrain the gas density in the high-velocity absorption components in our sample. A similar strategy was also used in many of our previous HVC absorption studies \citep[e.g.][]{Richter09,Richter13,Fox10,Fox13,Fox14}.

In our grid of Cloudy models, the absorbing clouds are assumed to be plane-parallel slabs illuminated on both sides by the local ($z=0$) UV background radiation from \citet{Haardt12}. For the modelling, we do not consider ionising radiation coming from stars in the Milky Way and in M31 due to the unknown location of the absorbers with respect to their stellar disks. As shown in the Milky Way radiation models by \citet{Fox05,Fox14}, the contribution of the local (stellar) radiation field in Milky-Way type galaxies is relevant for CGM absorbers within $\sim 100$ kpc of the plane, further depending on the position of the cloud with respect to the disk. The assumed flux of ionising photons in our Cloudy models therefore represents a lower limit. This needs to be kept in mind when it comes to the interpretation of the derived gas densities and thermal pressures from our models. We will further discuss these issues in Sect.\,5.1.3. 

The column densities for the above-listed ionisation states are modelled for different ionisation parameters and volume densities of hydrogen assuming a constant gas temperature of 10,000~K and relative solar abundances as given in~\citet{Grevesse10}. A temperature of $\sim 1-2\times 10^4$ K is typical for photoionised circumgalactic gas \citep[e.g.][]{Richter16} and similar temperatures have been derived in previous studies of individual HVCs  in the LG barycentre and anti-barycentre direction \citep{Collins05,Sembach99}. Since we do not have information on the H\,{\sc i} column density for most of the absorbers in our sample, we fix the overall metallicity of the gas to a value of $Z_X=0.1$ to indirectly constrain log $N$(H\,{\sc i}) from the observed ion column densities. A metallicity of $Z_X=0.1$ is characteristic for many of the most distant HVCs, such as the Magellanic Stream, Complex C, and Complex A \citep{Fox13,Richter01,Sembach04,Wakker99}.
 
\begin{figure}[htp]
\centering
\includegraphics[width=\hsize]{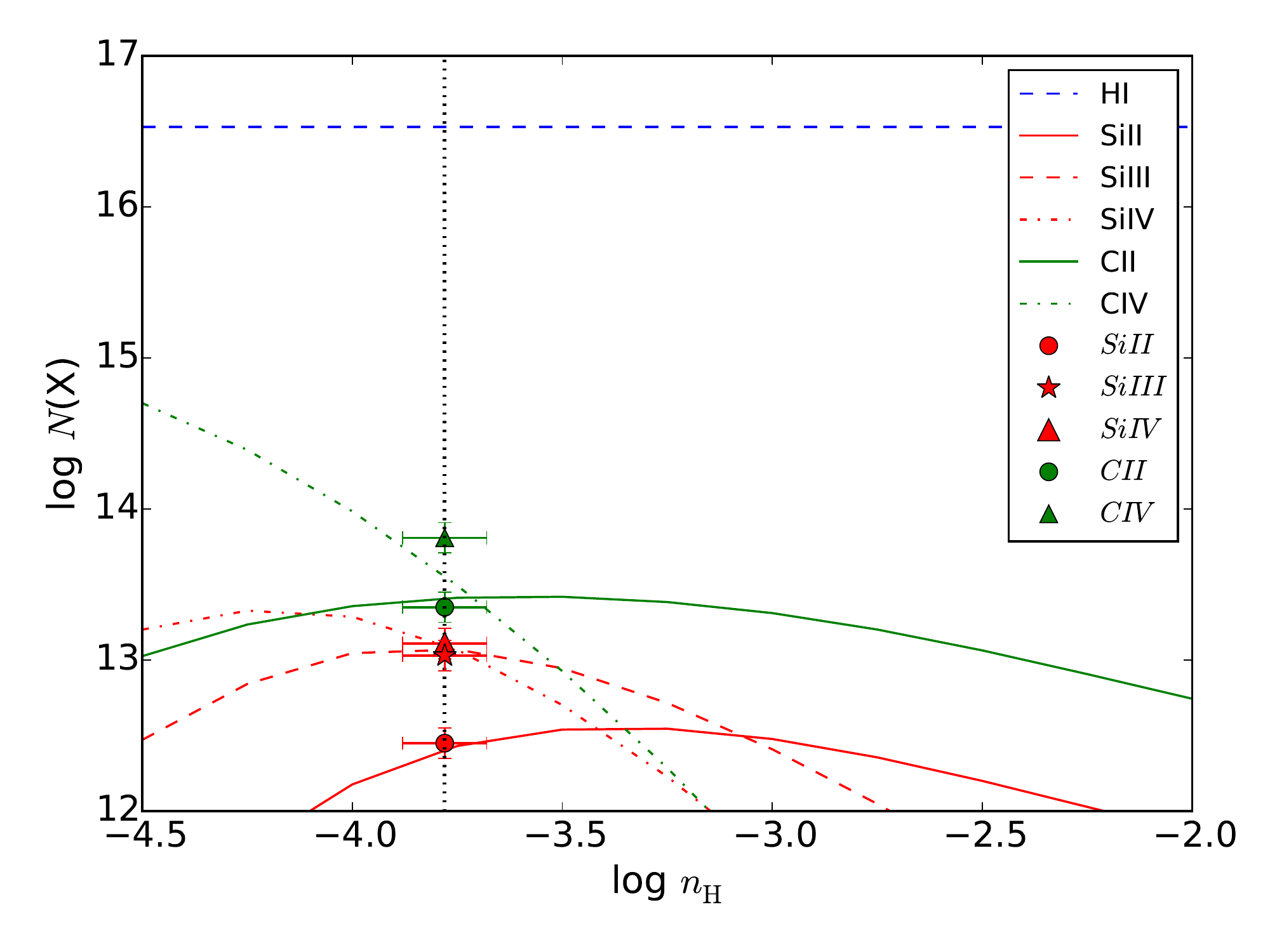}
\caption{Cloudy model for the HVC at $v_{\rm LSR}=-257$ km\,s$^{-1}$ in the direction of MR\,2251-178. Solid, dashed and dashed-dotted lines represent column densities predicted by the Cloudy model for different ions, while the symbols (dots, star and triangle) show the measured values. This model, with log $N$(H\,{\sc i})~=~16.53, fits Si\,{\sc ii}, Si\,{\sc iii}, Si\,{\sc iv}, and C\,{\sc ii} at log\,$n_{\rm H}=-3.78$, which is indicated by the black dotted line.}
\label{cloudy}
\end{figure}

For some HVCs in our sample, not all elemental abundances fit to the same Cloudy model, suggesting a multiphase-nature of these absorbers. A similar behaviour has also been found in CGM absorbers around other galaxies \citep[see detailed discussion in][]{Richter16}. More highly ionised species are likely to be found in the outer parts at lower hydrogen volume densities (log\,$n_{\rm H}$), while the less ionised species are found in the cloud cores at higher log\,$n_{\rm H}$. However, because not all multiphase clouds necessarily have this (rather oversimplified) internal structure and because Si is also affected by dust depletion effects, our strategy is to fit the Si and C ions independently and then evaluate possible discrepancies with respect to dust depletion and column-density errors for each absorber individually. Si\,{\sc iv} and C\,{\sc iv} are matched together only when Si\,{\sc ii}, Si\,{\sc iii}, and C\,{\sc ii} match the same model at the same log\,$n_{\rm H}$. For all models, the Si\,{\sc iii} is used as anchor-ion, as it is detected in all but one of the 54 HVCs (and for all HVCs in the anti-barycentre direction). The log\,$n_{\rm H}$ derived for the one HVC component without Si\,{\sc iii} is not considered, as the more highly ionised Si\,{\sc iv} and C\,{\sc iv} trace a different gas-phase. For two barycentre HVCs and one anti-barycentre HVC, Si\,{\sc iii} was the only securely detected ion, making it impossible to obtain log\,$n_{\rm H}$. For additional 15 HVCs (10 HVCs in the barycentre sample, 5 HVCs in the anti-barycentre direction) we did not find satisfying solutions in our Cloudy models (possibly related to blending and saturation problems), so these absorbers are not further considered in our subsequent discussion. In Fig.~\ref{cloudy} we show a typical example for our Cloudy modelling approach. A characteristic error of 0.1 dex is assigned to the column densities of the different ions and for log\,$n_{\rm H}$, this is in accordance with errors found in R17.

For this project, the ionising radiation field of \citet{Haardt12} is used. The intensity of this background field is lower than that of the \citet{Haardt01} background model and also lower than what is expected from recent measurements of the local HI photoionisation rate \citep{Kollmeier14,Wakker15,Shull15}. Generally, a lower intensity of the background field results in higher gas densities. Additionally, the \citet{Haardt12} spectrum is harder than the older version, leading to higher metallicities \citep[e.g.][]{Wotta19,Chen17}. To explore the uncertainties related to the scaling and shape of the UV background, we examplarily model the absorber towards MR 2251-178 adopting the \citet{Haardt01} background as it is implemented in Cloudy as table HM05. The change in the spectral energy distribution of the ionizing radiation leads to changes in the column density curves (see Fig.~\ref{cloudy}). For the HM05 spectrum, the peak of $\log N(\ion{Si}{iii})$ is closer to the curve of $\log N(\ion{C}{ii})$. For the MR 2251-178 example system, this results in a density of $\log n_{\mathrm{H}} = -3.55$ instead of $-3.78$, thus $0.23$ dex difference. While this difference is rather small, it should be noted that the choice of the ionizing radiation field does affect our results to a certain degree.


\begin{figure}[htp]
\centering
\includegraphics[width=0.5\textwidth]{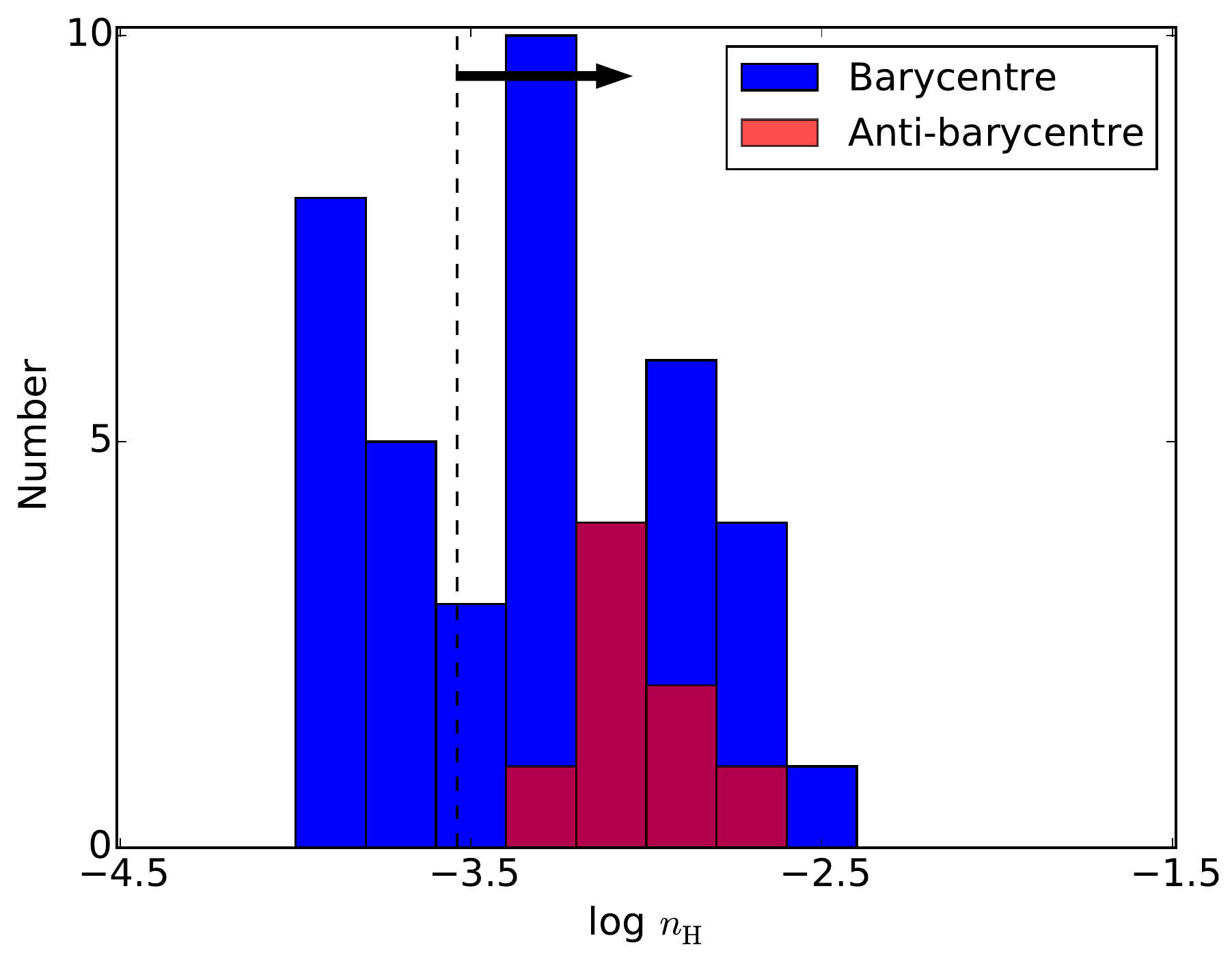}  
\caption{Histogram of logarithmic hydrogen volume densities for 41~HVCs in the barycentre direction (blue) and 8~HVCs in the anti-barycentre direction (red), based on our Cloudy modelling of these absorbers presented in Sect.\,4. The black, dashed line and the arrow indicate the region for which the Milky Way radiation field plays a role, see Sect.\,5.1.3.}
\label{hden}
\end{figure}


\section{Discussion}

\subsection{Distribution of gas density}

In Fig.~\ref{hden}, we show the distribution of the derived logarithmic gas densities, log $n_{\rm H}$, as a histogram. In the barycentre direction (blue colour), the densities are spread over a large range (log\,$n_{\rm H}=-2.5$ to $-4.0$), with more than 35 percent of the absorbers having very low densities log\,$n_{\rm H}\leq -3.5$. In the anti-barycentre direction the densities are confined to a much smaller range, log\,$n_{\rm H}=-2.7$ to $-3.3$.


\begin{figure*}[htp]
\centering
\includegraphics[width=0.9\textwidth]{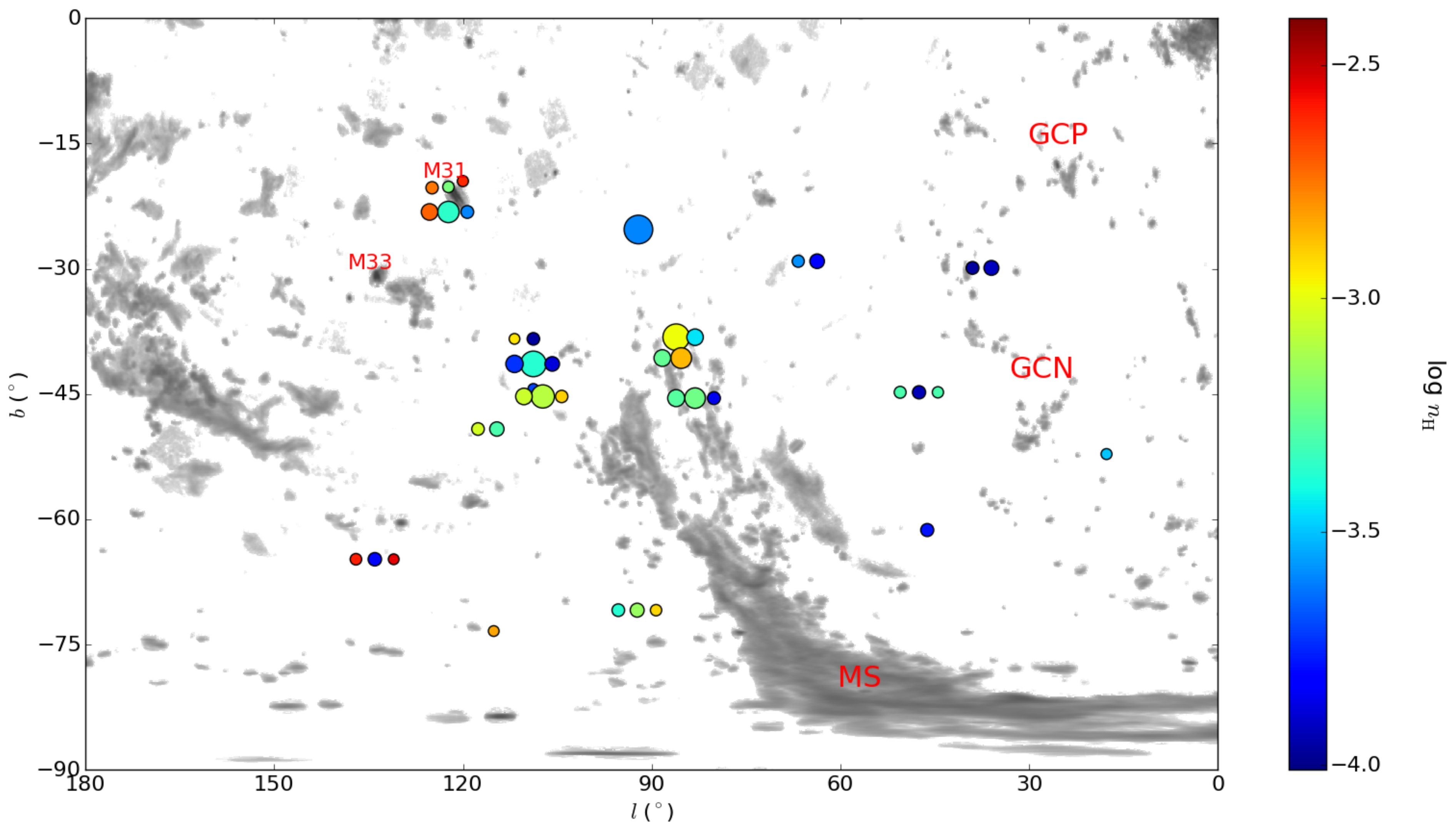}      
\caption{Sky-distribution of gas densities (colour coded; see right vertical bar) for 41 HVCs in the barycentre direction with log\,$n_{\rm H}$ measured from Si\,{\sc iii}, overlaid on the H\,{\sc i} map with data from \citet{Westmeier18}, which is based on the LAB survey \citep{Kalberla05,Kalberla07}. For displaying purposes, HVCs along the same sightlines have been given an angular offset. Size of the dots scales with the HVC's velocity. Important H\,{\sc i} complexes are indicated.
 }
 \label{hdenHI}
 \end{figure*}  


\subsubsection{HVCs in the barycentre direction}

Figure~\ref{hdenHI} shows the hydrogen volume densities for HVCs in the barycentre direction plotted on top of the H\,{\sc i} 21cm distribution of high-velocity gas in this region of the sky (grey-shaded areas). There are more HVCs with higher log\,$n_{\rm H}$ near the MS and near M31 than in the direction of the Galactic Centre Negative (GCN) Complex. Distance estimates for Complex GCN range from 15 to 35~kpc, based on computations of a hypothetical orbit that the clouds composing Complex GCN might follow \citep{Jin10}. If Complex GCN is connected to the Leading Arm of the Magellanic System, as suggested by simulations from \citet{Diaz11}, the distances would be between 14 to 22~kpc. However, Complex GCN rather might be a conglomerate of scattered gas clouds with different origins and distances, as pointed out by \citet{Winkel11}. Indeed, the absorption properties of the GCN absorbers towards Mrk\,509 and PKS\,2155$-$304 \citep{Sembach99} indicate extremely low gas pressures of $P/k=1-5$ cm$^{-3}$ K, inconsistent with gas located in the inner ($d<50$ kpc) Milky Way halo, but rather pointing toward a LG origin for these gas clouds.

Some of the highest densities are found in the absorbers in the general direction of M31 (see Fig.~\ref{hdenHI}). These have relatively low velocities, suggesting that they are  close to the MW disk and not associated with M31 itself. Absorbers near the tip of the MS show the highest velocities. In this region, the MS has velocities in 21cm between $\sim-250$ to $-350$ km\,s$^{-1}$ \citep{Kalberla05}, demonstrating that---from the kinematics---those clouds could well be associated with the MS. However, the absorber at ($l$,$b$) $\sim (110,-40$) has $v_{\rm LSR}\sim-400$ km\,s$^{-1}$, indicating it is not part of the MS. The low densities (log\,$n_{\rm H}\sim -3.9$) of some absorbers in this region are also indicative of an extragalactic origin. We will further discuss these aspects in Sec.\,5.3, where we include the pressure estimates as additional constraints for the origin of the high-velocity absorbers in the barycentre direction.

\subsubsection{HVCs in direction of the anti-barycentre}

Since in this region there are no prominent clouds seen in the H\,{\sc i} 21cm HVC surveys, we do not provide a density map as for the barycentre direction.  The sightlines are all relatively close together on the sky and in velocity (Fig.~\ref{skydist}) and the spread in log\,$n_{\rm H}$ is much smaller than for the anti-barycentre sample (Fig.~\ref{hden}). This coherency points at a common origin of these high-velocity absorbers as part of a coherent, ionised high-velocity gas structure located in the inner or outer Milky Way halo.

\subsubsection{The role of the Milky-Way radiation field}

As mentioned above, our ionisation modelling is based on the assumption that the HVC absorbers are exposed only to the extragalactic radiation field at $z=0$, i.e., any contribution from the Milky Way radiation field is ignored so far. Since for a given set of input column densities of various metal ions Cloudy delivers the ionisation parameter $U$ as primary result, there is a degeneracy between the gas density and the intensity of the ionising field in the interpretation of $U$, {\it if} Galactic ionising photons are important in addition to the UV background field. That means, the same value for $U$ could be achieved for different combinations of $n_{\gamma}$ and $n_{\rm H}$ at different locations inside and outside the Milky Way halo.

\citet{Fox05, Fox14} have modelled in detail the flux of Galactic ionising radiation, demonstrating that the Milky Way radiation field becomes important for galactocentric distances $r\leq 100$ kpc. The exact strength of the local radiation field in the Milky Way halo depends, however, strongly on location relative to the Galactic disk. Since we do not know the location (and distance) of the HVC absorbers in our sample with respect to the disk, the influence of the Milky Way radiation field on the derived HVC gas densities represents a critical aspect in the interpretation of our Cloudy models. To overcome the degeneracy between gas density and photon density we make use of the fact that the gas density in photoionised gas clouds at $T=10,000-20,000$ K situated in a galaxy potential well is constrained by the local gas {\it pressure}, which is expected to decline rapidly with increasing galactocentric distance. For a fully ionised gas containing 10 percent helium (in number), the thermal gas pressure, $P$, is given by the relation $P/k=2.3\,n_{\rm H}T$, where $k$ is the Boltzmann constant, $n_{\rm H}$ the hydrogen volume density, and $T$ the temperature of the gas. From the coronal gas model by \citet{Miller15} follows that $P/k =Cr^{-1.58}$ or $n_{\rm H}=Dr^{-1.58}$, where $C$ and $D$ are constants with values of $C=3.31\,10^4$ K\,cm$^{-1.42}$, $D=0.435\,C\,T^{-1}$, and $r$ is the galactocentric distance. By combining the density constraints from the \citet{Fox06, Fox14} radiation field model and the \citet{Miller15} pressure relation we are able to define a threshold value for $U$ that separates absorbers which potentially are influenced by the Galactic radiation field (thus being located at $r\leq 100$ kpc) from absorbers that must be located at $r>100$ kpc (i.e., absorbers for which no radiation-field/pressure model solution exists at $r\leq 100$ kpc). Taking the extragalactic radiation field as reference, this threshold in $U$ corresponds to a density threshold of log $n_{\rm H}=-3.54$.

We conclude that all absorbers, for which our initial Cloudy model delivers gas densities log $n_{\rm H}\leq -3.54$, must be located beyond $r= 100$ kpc and thus cannot be significantly influenced by radiation from the Milky Way disk. For all other absorbers with log $n_{\rm H}>-3.54$, we do not know where they are located in the Milky Way halo. They could be either inside or outside $r= 100$ kpc and therefore the gas densities derived for them from our initial UV background Cloudy model must be regarded as lower limits.

\subsection{Density-velocity plane}

A correlation is expected between $v_{\rm LSR}$ and log\,$n_{\rm H}$ if high velocity and low density are both signs of clouds being at larger distance. In fact, halo clouds are expected to slow down as they get closer to the Galactic disc because the pressure and density of their surrounding medium increases and so does the effect of ram pressure, which scales with $\rho v_{\rm infall}^2$. As a consequence, clouds in the lower halo are expected to have relatively low velocities, while clouds in the outer halo,  or even beyond, can have much higher velocities. Since only the radial component of the space velocity is observed, however, there might be a population of distant clouds that move with large tangential velocities through the outer halo, while we see them as low-velocity HVCs (or even IVCs) because of their small radial velocity component. Clearly, a large sample of Milky Way halo absorbers is required in order to identify possible trends between $v_{\rm LSR}$ and log\,$n_{\rm H}$.

Note that the velocities from the HVCs in the direction of M31 are mostly in the range $-250 \lesssim v_{\rm LSR}\lesssim -120$ km\,s$^{-1}$, with two HVCs having absolute velocities $>$ 300 km\,s$^{-1}$. The systemic velocity of M31 is about $-300$ km\,s$^{-1}$ and HVCs with $-300\lesssim v_{\rm LSR} \lesssim -150$ km\,s$^{-1}$ might belong to either M31 or its dwarf companions \citep{Lehner15}. Thus, some of the HVCs in this direction could be part of the CGM of M31 instead of that of the Milky Way.

In the upper panel of Fig.~\ref{vbinhden}, we plot log\,$n_{\rm H}$ against $v_{\rm LSR}$ for all 49 absorption components. In the lower panel of Fig.~\ref{vbinhden}, we bin the data points in 50 km\,s$^{-1}$-wide velocity bins, where we regard the individual densities as fixed values (i.e., in this plot we ignore that densities may represent lower limits).

\subsubsection{HVCs in direction of the barycentre}

For the HVCs in the direction of the LG barycentre, the upper panel of Fig.~\ref{vbinhden} indicates a large scatter in the $v_{\rm LSR}$ /log\,$n_{\rm H}$ parameter plane. This trend is expected from the large range that the barycentre absorbers cover in velocity (Fig.~\ref{vdist}) and density (Fig.~\ref{hden}). The sample appears to split into two populations: low-density absorbers with log\,$n_{\rm H}\leq -3.54$ that appear to cluster near $v_{\rm LSR}=-300$ km\,s$^{-1}$, and absorbers with high-densities (or density limits) that uniformly scatter in the region log\,$n_{\rm H}>-3.54$ and $v_{\rm LSR}=-100$ to $-400$ km\,s$^{-1}$. The large scatter in log\,$n_{\rm H}$ is also reflected in the large error bars seen in the binned version (Fig.~\ref{vbinhden}, lower panel), which indicate the standard deviation for in log\,$n_{\rm H}$ in each bin. For the barycentre sample, the bin with $v_{\rm LSR}\geq-100$ km\,s$^{-1}$ has only one HVC, so an error of $0.1$ dex is assumed. The same is true for the last bin ($+250<v_{\rm LSR}<+300$~km\,s$^{-1}$) in the anti-barycentre sample. Standard deviations in the other two bins in the anti-barycentre sample are smaller than the size of the symbols. Despite the large scatter, the binned version shows a mild trend, in which the average density increases with decreasing absolute radial velocity. As mentioned above, this behaviour is expected in the scenario, in which distant clouds move through the Milky Way halo with higher velocities than nearby clouds, the latter being slowed down by ram-pressure forces that act predominantly in the inner halo. While this trend is plausible, we refrain from providing a deeper discussion on this issue at this point to avoid an over-interpretation of the data. The large scatter in log\,$n_{\rm H}$ in the overall sample and the fact that for log\,$n_{\rm H}\leq -3.54$ our Cloudy modelling provides only lower limits if the Galactic radiation field is taken into account (see Sect.\,4) may also be (at least partly) responsible for this trend, but the influence of these effects on the observed $v_{\rm LSR}/$log\,$n_{\rm H}$ distribution remains unknown. 

The Lee statistics \citep{Lee79} can be used to determine if there are indeed two distinct populations in the density distribution of our absorber sample and to provide an estimate for the characteristic density that separates these two populations from each other (separation value). The first step is to split the sample into two subsamples for different values of  log\,$n_{\rm H}$. Next, the second moment of the total population is multiplied by the total number of data points which is then divided by the sum of values derived on either side of the separation value, resulting in the Lee parameter, $L$. The value of $L$ will be at its maximum for the most likely separation value (see \citet{Fitchett88} for more details). To evaluate $L$ for our absorber sample, all limits on the gas density are regarded as absolute values. The maximum $L$ comes out to $2.45$ and is found at a density of log\,$n_{\rm H} = -3.44$, while the minimum $L$ is $0.09$ at log\,$n_{\rm H} = -2.44$. The split at log\,$n_{\rm H} = -3.44$ is marginally higher than the value derived above ($-3.54$), resulting in two more absorbers being assigned to the low-density subsample. Overall, however, the results from the Lee statistics agree very well with the results based on the estimated influence of the Galactic radiation field (Sect.\,5.1.3).

Despite the uncertainties in modelled densities, the existence of a distinct population of barycentre absorbers with log\,$n_{\rm H}\leq -3.54$ and $v_{\rm LSR}\leq~-120$ km\,s$^{-1}$ is securely constrained by the distribution shown in Fig.~\ref{vbinhden} and further strengthened by the Lee statistics. These results point toward photoionised halo absorbers with very low gas pressures (see subsequent discussion in Sect.\,5.3).

\subsubsection{HVCs in direction of the anti-barycentre}

\begin{figure}[htp]
\centering
\includegraphics[width=\hsize]{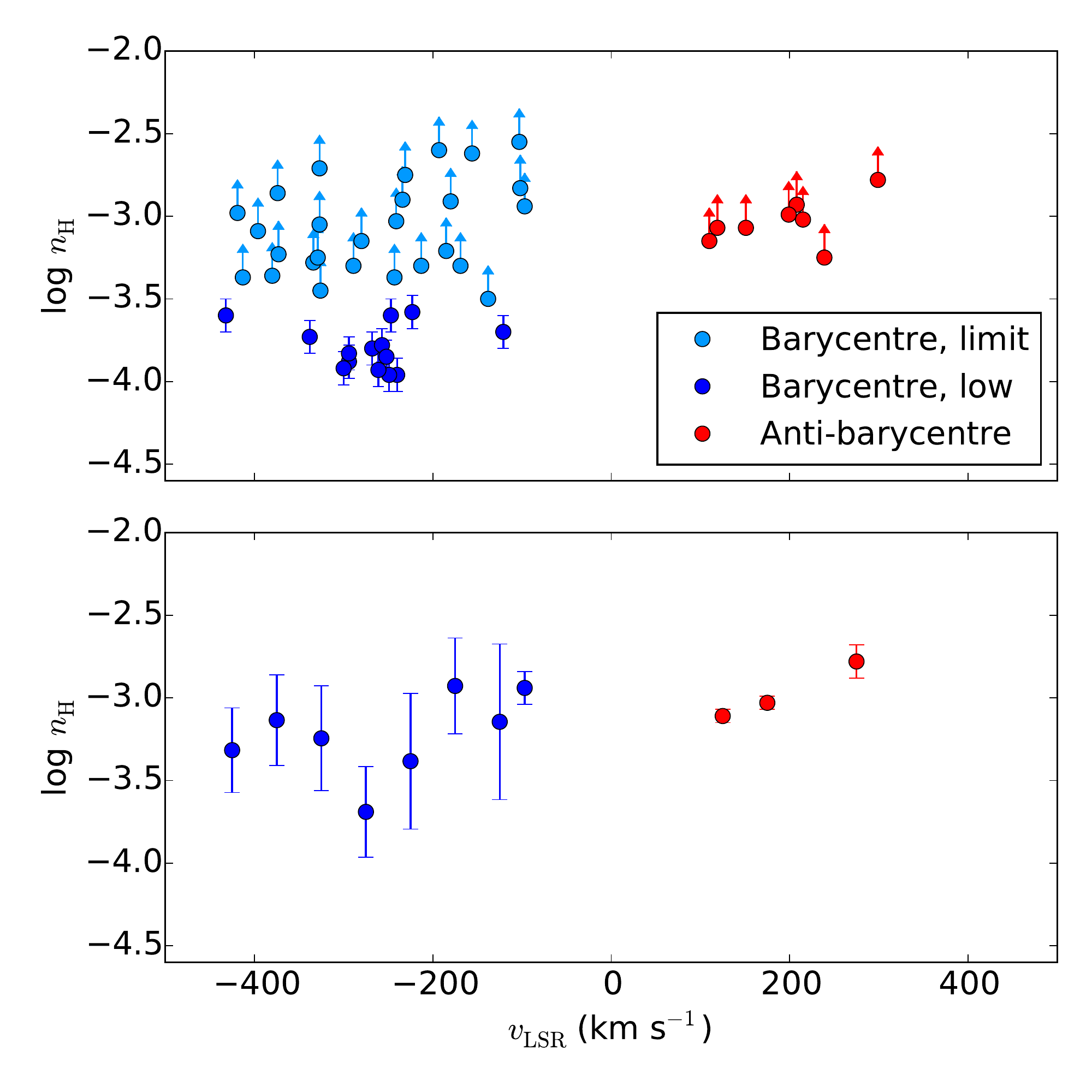}
\caption{Volume density of hydrogen plotted against $v_{\rm LSR}$ for our HVC absorber sample. The blue data points show the barycentre absorbers, the red data points indicate the anti-barycentre absorbers. The densities are taken from our Cloudy modelling. {\it Upper panel:} data points for the 49~HVCs with reliable $n_{\rm H}$ measurements are shown. Lower limits are indicated by light blue symbols and marked with the arrow symbol. HVCs with log\,$n_{\rm H}~\leq~-3.54$ are shown in dark blue. {\it Lower panel:} same as above, but data are binned in 50 km\,s$^{-1}$-wide velocity bins. For this plot, we regard the individual densities as fixed values (i.e., in this plot we ignore that densities may represent lower limits)}
\label{vbinhden}
\end{figure}

For the HVCs in the anti-barycentre direction, all absorbers cluster in the velocity/density plane in the region $v_{\rm LSR}\sim+300$~km\,s$^{-1}$ and log\,$n_{\rm H}\sim-3.0$ with very little scatter (Fig.~\ref{vbinhden}, upper and lower panel).  Thus, the anti-barycentre absorbers have -- on average -- significantly higher densities and lower absolute radial velocities than the barycentre absorbers. This striking asymmetry between barycentre/anti-barycentre HVCs has also been noted and discussed in R17. 

It should be noted that the anti-barycentre sample contains only 8 absorbers with well-constrained gas densities, compared to the 41 absorbers in the barycentre sample. Only $\sim$~35\% of the barycentre absorbers have low log\,$n_{\rm H}<-3.54$.

If the anti-barycentre sample would have the same fraction of low density clouds than the barycentre sample ($\sim$~35\% having log\,$n_{\rm H}<-3.54$), one would expect to find only 3 of these 
systems in the anti-barycenter direction at this sample size.
Because of this low-number statistics, we could have missed low density absorbers in the anti-barycentre direction that are not covered by our sample.


\subsection{Gas pressures and galactocentric distances}

HVCs are expected to be in pressure equilibrium with their environment, or else they would not exist very long. Since the coronal gas pressure in the inner Galactic halo ($r<50$ kpc) is expected to be $P/k~\gtrsim 100$ K~cm$^{-3}$ \citep{Wolfire95,Miller15}, HVCs in the inner Galactic halo should have similar pressures. We here assume a constant temperature of $10,000$~K for the photoionised HVCs. For the errors in the pressure estimates we allow the temperature vary in the range $T=5,000-20,000$ K and also consider the individual errors derived for the densities from the Cloudy modelling.

\begin{figure}[htp]
\centering
\includegraphics[width=\hsize]{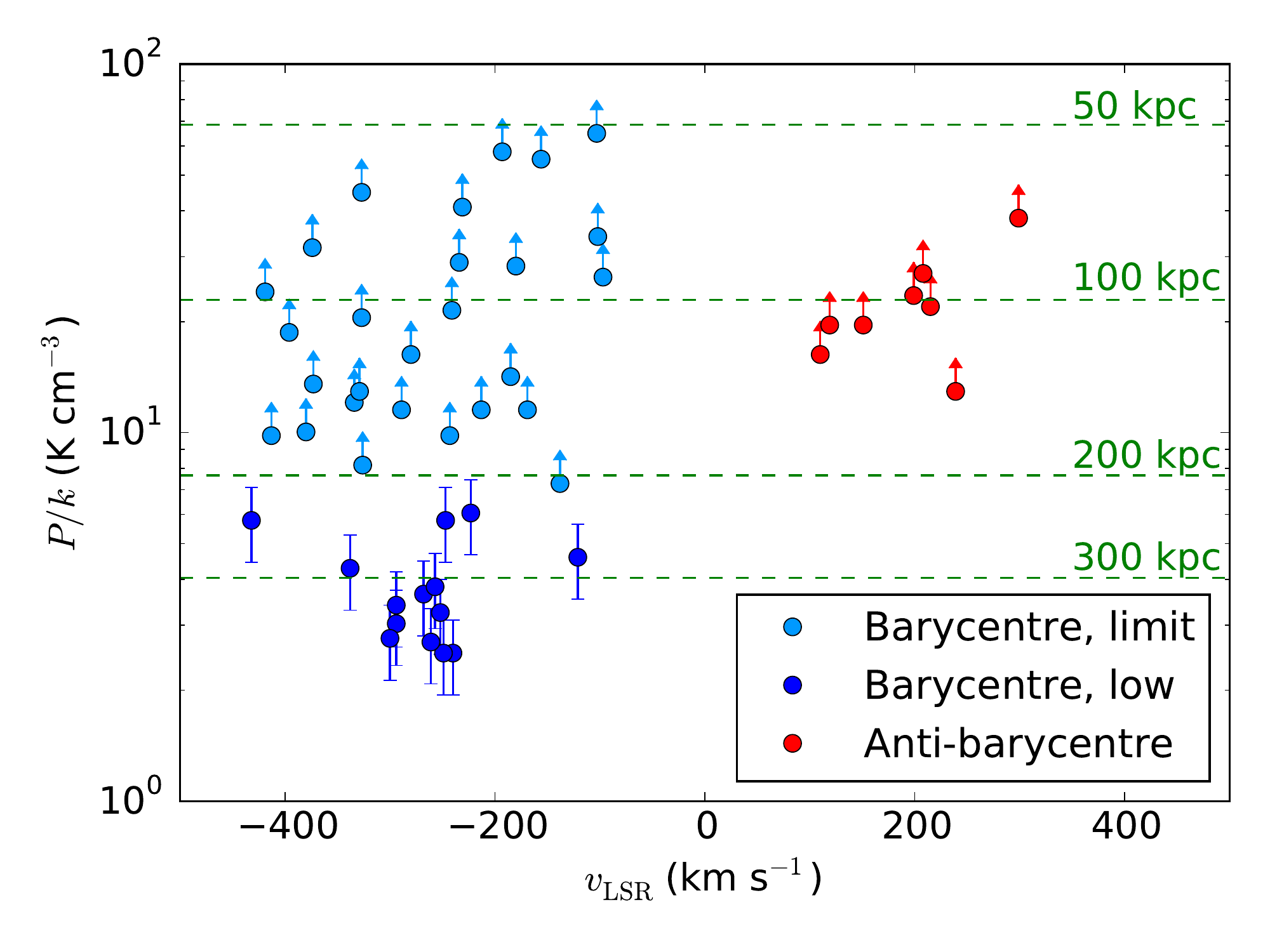}
\caption{Same as Fig.~\ref{vbinhden}, but now with the pressure values, log $P/k$, on the y-axis. Lower pressure limits are indicated with the arrows (see also Sect.\,5.2.1). The dashed green lines show distance-scale based on the halo pressure model from \citet{Miller15}. 
}
\label{Pk}
\end{figure}

Looking at the range of hydrogen densities (or limits) derived for the barycentre and anti-barycentre sample, all HVCs have pressure values (or lower limits) in the range $P/k \sim 2.5-50$ K\,cm$^{-3}$, as indicated in Fig.~\ref{Pk}. For the 41 absorbers in the barycentre direction, for which absolute values of log\,$n_{\rm H}$ could be determined (see Sect.\,5.1.1), we derive gas pressures in the range $P/k\sim 2.5-10$ K\,cm$^{-3}$, which is far lower than what is expected for the inner halo. The lowest pressures are found for the HVCs in the direction of Complex GCN and the HVCs that are (in projection) close to the tip of the MS. 

Figure~\ref{Pk} shows $v_{\rm LSR}$ plotted against $P/k$ together with a distance scale (dashed green lines), based on the pressure-distance relation described in \citet[][see their Sect.\,5.4]{Miller15}. Applying their model to our data, we calculate for each absorber the galactocentric distance (or limit) that corresponds to the derived pressure (or limit). As can be seen, the distances derived for the low-density barycentre HVCs with log\,$n_{\rm H}\leq -3.54$ are very large, $r>200$ kpc, thus beyond the assumed virial radius of the Milky Way. 

Note that the \citet{Miller15} pressure model considers only the MW gravitational potential, but not the LG potential, which defines the pressure floor for any gas absorber located inside the LG. 
The pressure has been estimated to decrease from $P/k \sim 10$ K\,cm$^{-3}$, at a distance of 200 kpc from the Milky Way centre, to  $P/k \sim 0.1$ K\,cm$^{-3}$ in the LG intra-group medium in the direction of M31 \citep{Nuza14,Richter17}. Therefore, the lowest pressure values derived from our sample correspond to gas outside of the Milky Way's virial radius possibly residing in the LG's intra-group medium.

The low pressures derived here can be compared to CGM clouds within the virial radius of other, more distant galaxies. Gas pressures of clouds in the CGM of low-redshift galaxies have been estimated by \citet{Werk14} and \citet{Stocke13}, indicating $P/k \sim 3 - 100$ K\,cm$^{-3}$ within the virial radius. Thus, the lowest pressures in our HVC sample fall just below the range suggested for CGM gas in the local Universe.


\subsubsection{Two separate HVC populations}

From our pressure modelling we conclude that all barycentre HVC absorbers with gas densities lower than log\,$n_{\rm H}=-3.54$ (14~absorbers) and that are predominantly associated with Complex GCN and the outer boundary of the MS, most likely are located beyond the Milky Way's virial radius in the in the LG. For the other, remaining barycentre and anti-barycentre absorbers with log\,$n_{\rm H}> -3.54$ (35 absorbers) the derived density and pressure limits do not provide constraints on the location of these absorbers inside or outside the Milky Way halo. 

The barycentre HVCs near Complex GCN and the tip of the MS  represent a distinct population of high-velocity absorbers that are peculiar also in other properties. As discussed in R17, the region near Complex GCN and the tip of the MS contains very little H\,{\sc i} 21cm emission, but is very pronounced in O\,{\sc vi} absorption \citep{Sembach03,Wakker03}. In addition, the many scattered 21cm clumps do not show the typical two-phase characteristics of nearby HVCs with cold cloud cores and extended diffuse gas layers, indicating that they reside in a very low-density environment. 

With only the 14 barycentre absorbers having distinct properties that point to a LG origin, our study does not provide further evidence for the scenario, in which the observed velocity dipole of HVC absorbers in the barycentre and anti-barycentre absorbers is a result of the overall motion of the Milky Way with respect to the Local Group's intragroup medium (R17). Since the distances found in our study are derived from the gas densities and pressures, they depend - among other assumptions - on the background radiation field and the assumed Milky Way halo model. The low-number statistics for the anti-barycentre sample adds to these uncertainties. However, our study also does not exclude such a model and the observed large negative velocities of the LG absorbers in the barycentre direction may well be related to the Milky Way's overall motion towards M31 and its companions \citep[see also][]{Nuza14}.

\subsection{Masses and sizes of the HVCs}

\begin{figure}[htp]
\centering
\includegraphics[width=\hsize]{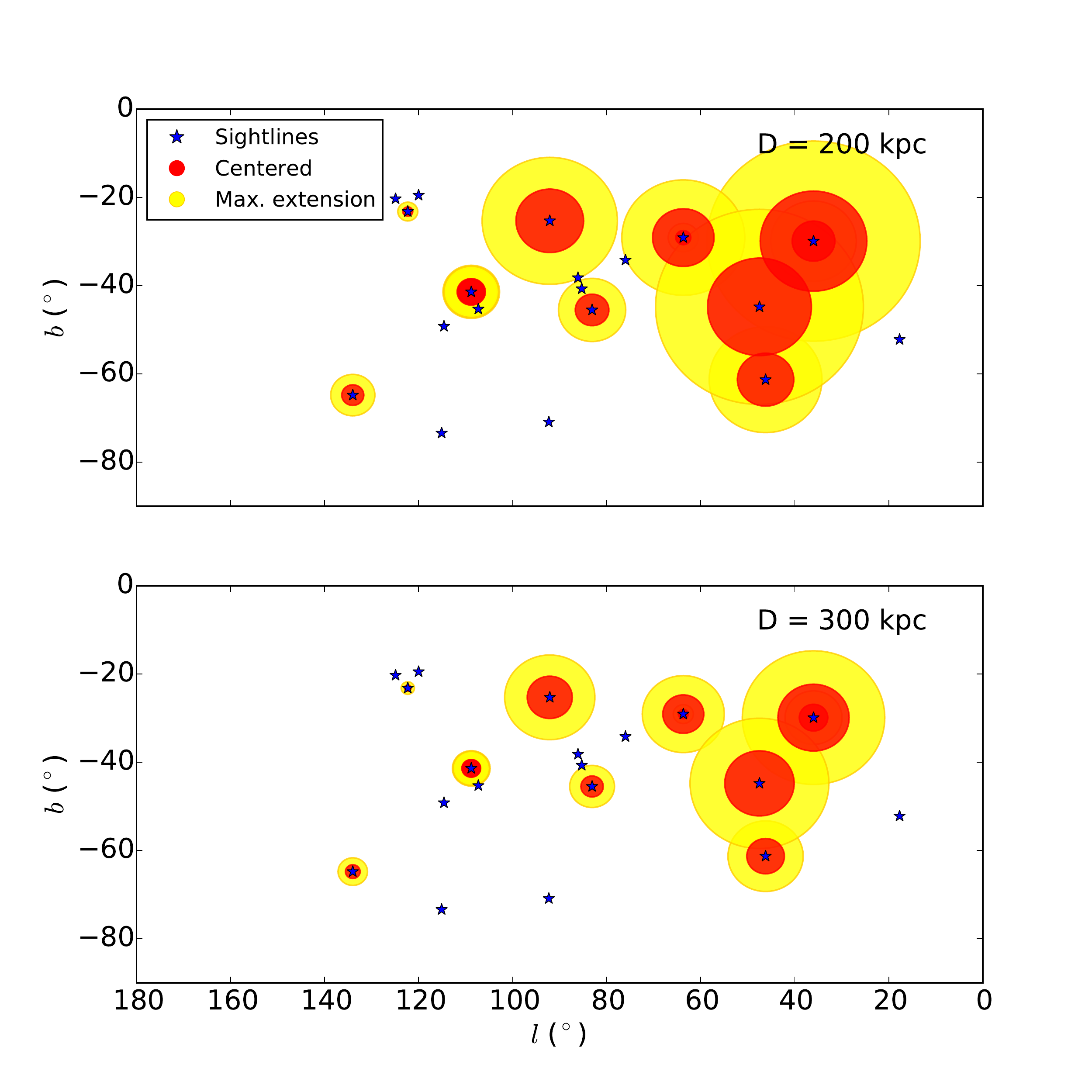}
\caption{Positions and angular sizes on the sky of the absorbers with log\,$n_{\rm H} \leq -3.54$ in the direction of the LG barycentre. For sightlines with multiple low-density absorbers, the one with the largest angular size has been plotted. Red filled circles indicate the size if the cloud would be centered on the sightline along which it is detected. Yellow circles instead indicate the maximum extent of the cloud, assuming that the QSO sightline passes the edge of the cloud.}
\label{angular}
\end{figure}

The sizes and masses of the absorbers can be estimated from log\,$N$($\rm H$) and log\,$n_{\rm H}$, based on the values derived from the Cloudy models. Note that a direct measurement of log\,$N$($\rm H$\,{\sc i}) from Ly$\alpha$ absorption is not possible, as Galactic disk emission and absorption dominates the Ly$\alpha$ spectrum up to about $400$~km~s$^{-1}$. We here compute sizes and masses for the subsample of barycentre clouds with log\,$n_{\rm H} \leq -3.54$ by using the Cloudy results in combination with equations (6)-(8) presented in \citet{Richter09}. From this we derive linear diameters  between $5$ and $39$ kpc, while the masses range from $\sim 10^{5}$ to $10^{8}$ M$_\odot$ for the most massive cloud, with the median mass being at $4.0 \cdot 10^6$ M$_\odot$. The angular sizes of the absorbers on the sky depend on their distances. Assuming a distance of 200 kpc, the estimated angular sizes are up to 11 degrees. Assuming instead that the absorbers are located at 300 kpc reduces their angular sizes by a factor of $2/3$. Figure~\ref{angular} displays the location of the barycentre absorbers on the sky together with their angular sizes.

If we further assume that the absorbers are centered on the QSO sightline along which they are observed, none of the absorbers are expected to cover multiple QSO sightlines. This is true for both 200 kpc and 300 kpc distances, although there are two systems near the tip of the MS where the estimated extent of the clouds almost overlaps with other QSO sightlines. Since those other sightlines do not show low-density absorption, one could conclude that these clouds cannot be at distances closer than 200 kpc, as they would be picked up by multiple QSO sightlines. This conclusion depends, however, on various assumptions, for instance the assumed spherical geometry of the clouds.

If the absorbers are {\it not} centered on the sightline along which which they are detected, there are several cases where clouds may cover multiple sightlines. However, only in the case of the four absorbers in the direction of Complex GCN (see also Fig.~\ref{hdenHI}) do the different sightlines all show coherent low-density absorption that could be associated with overlapping absorbers. As seen in Fig.~\ref{hdenHI}, these possibly overlapping absorbers in the direction of the Complex GCN all have similar velocities and similar densities. If this is indeed one and the same low-density cloud stretching over multiple QSO sightlines, a distance of about 200 kpc is needed, since there is no more overlap at 300 kpc. However, a conglomerate of small clouds with similar properties could also mimic such an absorption pattern. This would fit the picture of the H\,{\sc i} gas in this region, as seen in Fig.~\ref{hdenHI}.

In conclusion, the estimated masses and sizes 
do not provide any additional firm constraints on the distance and origin of the barycentre absorbers, but they are fully in line with a LG origin of these clouds, as concluded from the low gas pressures.

\section{Summary \& Conclusions}

In this paper, 68 high-velocity absorption systems in a sample of 29 HST/COS spectra were studied to learn more about the origin of these HVCs and their relation to the intragroup medium of the LG.

Of these spectra, 19 are in the direction of the LG barycentre and 10 are in the LG anti-barycentre direction. The velocity range studied for the barycentre sample is $-100<v_{\rm LSR}<-400$ km\,s$^{-1}$, while that for the anti-barycentre sample is $+100<v_{\rm LSR}<+400$ km\,s$^{-1}$. We measured column densities for the ions Si\,{\sc ii}, Si\,{\sc iii}, Si\,{\sc iv}, C\,{\sc ii} and C\,{\sc iv} in these absorbers by modelling in detail the absorption patterns with synthetic spectra. These column densities were then used together with Cloudy to model the ionisation conditions in the absorbers and to determine the local gas densities, log\,$n_{\rm H}$.

The density constraints that we derived from this modelling indicate that the absorbers span a large range in gas densities from log\,$n_{\rm H}=-2.5$ to $-4.0$. The positive-velocity absorbers all have log $n_{\rm H}>-3.3$, corresponding to thermal pressures of $P/k> 11.5$ K\,cm$^{-3}$ for $T=10,000$ K. These absorbers must be located within $R_{\rm vir}=200$ kpc of the Milky Way in the inner and outer halo. The negative-velocity absorbers are split into two different populations. One population (27 systems) has log $n_{\rm H}>-3.54$ and $P/k> 7.3$ K\,cm$^{-3}$, thus similar to the positive-velocity absorbers, indicating a location within $R_{\rm vir}$. The other population (14 systems) consists of absorbers with very low gas densities (log\,$n_{\rm H} \leq -3.54$) and very low thermal pressures ($P/k<7.3$ K\,cm$^{-3}$), suggesting that these absorbers trace gas situated outside the Milky Way halo in the LG intragroup medium. The derived gas densities and pressures depend on several assumptions, such as the ionising radiation field. Therefore, the derived distances of the absorbers conservatively should be regarded as approximate values. The low-density, low-pressure absorbers appear to be spatially and kinematically connected to HVC Complex GCN and the tip of the Magellanic Stream, but also could lie behind these complexes. A LG origin for these absorbers is further supported by other observed peculiarities, such as the lack of a pronounced core-envelope structure \citep{Winkel11} and the excess of O\,{\sc vi} absorption \citep{Sembach03}.

To more precisely pinpoint the exact location of the (predominantly ionised) low-density LG absorbers beyond the virial radius of the Milky Way using the bracketing method \citep[e.g.][]{Wakker07,Wakker08,Richter15}, combined spectroscopic observations of the most distant globular clusters and adjacent QSO sightlines would be highly desired. While the bracketing method can be used for nearby clouds using bright halo stars and QSOs \citep[e.g.][]{Peek}, this is not possible for distances near or outside the virial radius. The strategy of using distant globular clusters has to await future UV instrumentation, however, as the eligible background sources are too faint to be observed with HST/COS at realistic integration times. 

\begin{acknowledgements}
We would like to thank the referee for very useful comments and suggestions.
\end{acknowledgements}

\bibliographystyle{aa} 
\bibliography{references}

\clearpage
\begin{appendix}
\onecolumn

\setcounter{table}{0}
\renewcommand{\thetable}{A\arabic{table}}
\section{Chemical abundances}

\tabcolsep=0.11cm
\begin{longtable}{ l l l l l l l l l l }

  \caption{\label{tab:abundances} Chemical abundances}\\
  \hline\hline
   QSO Name & $l$ (\degree) & $b$ (\degree) & $v_{\rm LSR}$ (km s$^{-1}$)  & log\,$N$(Si\,{\sc ii}) & log\,$N$(Si\,{\sc iii}) & log\,$N$(Si\,{\sc iv}) & log\,$N$(C\,{\sc ii}) & log\,$N$(C\,{\sc iv}) & log\,$n_{\rm H}$\\
  \hline
  \endfirsthead
  \caption{continued}\\
  \hline\hline
   QSO Name & $l$ (\degree) & $b$ (\degree) & $v_{\rm LSR}$ (km s$^{-1}$)  & log\,$N$(Si\,{\sc ii}) & log\,$N$(Si\,{\sc iii}) & log\,$N$(Si\,{\sc iv}) & log\,$N$(C\,{\sc ii}) & log\,$N$(C\,{\sc iv}) & log\,$n_{\rm H}$\\
  \hline
  \endhead
  \hline
  \endfoot

2MASX-J01013113+4229356 & 124.9 & -20.3 & -231 & 13.71 & 13.40 & 13.21 & 14.59 & 13.50 & -2.75 \\
  &   &   & -193 & 13.24 & 12.69 & 13.04 & 14.52 & 13.90 & -3.80 \\
AKN-564 & 92.1 & -25.3 & -432 & 12.79 & 13.39 & 13.21 & 13.90 & 13.40 & -3.60 \\
HS-0033+4300 & 120.0 & -19.5 & -202 & ... & 13.15 & 13.18 & 14.03 & 13.98 & -3.40 \\
  &   &   & -156 & 13.32 & 12.85 & ... & 14.25 & 13.20 & -2.62 \\
  &   &   & -119 & 13.78 & 13.05 & ... & 14.86 & ...  & -2.30 \\
LBQS-0107-0235 & 134.0 & -64.8 & -268 & ... & 12.63 & 12.67 & 13.20 & 13.65 & -3.80 \\
  &   &   & -193 & 13.02 & 12.56 & ... & 13.27 & 13.00 & -2.60 \\
  &   &   & -103 & 13.14 & 12.66 & ... & 13.76 & 13.44 & -2.55 \\
MR2251-178 & 46.2 & -61.3 & -257 & 12.45 & 13.03 & 13.11 & 13.35 & 13.81 & -3.78 \\
MRK335 & 108.8 & -41.4 & -413 & 12.18 & 12.49 & ... & 13.32 & ...  & -3.37 \\
  &   &   & -338& 12.19 & 12.80 & ... & 13.34 & 13.28 & -3.73 \\
  &   &   & -294 & ... & 12.58 & ... & ... & 13.29 & -3.88 \\
  &   &   & -240 & ... & 12.37 & ... & ... & 13.27 & -3.96 \\
  &   &   & -121 & 11.98 & 12.54 & ... & 13.50 & ...  & -3.70 \\
  &   &   & -97 & 13.28 & 13.18 & ... & 14.16 & ...  & -2.94 \\
MRK-509 & 36.0 & -29.9 & -300 & 12.15 & 13.20 & 13.10 & 13.58 & 14.07 & -3.92 \\
  &   &   & -249 & ... & 12.56 & ... & 12.93 & 13.49 & -3.96 \\
MRK1513 & 63.7 & -29.1 & -294 & ... & 13.01 & 13.09 & 13.39 & 14.02 & -3.83 \\
  &   &   & -223 & ... & 12.76 & ... & 13.22 & 12.98 & -3.58 \\
NGC-7469 & 83.1 & -45.5 & -373 & 13.04 & 13.20 & 12.66 & 13.97 & ...  & -3.23 \\
  &   &   & -334 & 13.45 & 13.62 & 13.21 & 14.12 & ...  & -3.28 \\
  &   &   & -252 & ... & 12.70 & 12.80 & 13.23 & ...  & -3.85 \\
  &   &   & -185 & ... & 12.40 & 12.85 & 13.12 & ...  & -4.18 \\
PG0003+158 & 107.3 & -45.3 & -431 & ... & 12.40 & 13.06 & 13.12 & ... & -4.40 \\
  &   &   & -396 & 12.76 & 12.78 & ... & 13.70 & 13.35 & -3.09 \\
  &   &   & -327 & 13.25 & 13.19 & ... & 14.10 & ...  & -3.05 \\
  &   &   & -234 & ... & 12.53 & ... & 13.53 & 13.39 & -2.90 \\
  &   &   & -360 & ... & ... & 12.78 & ... & 13.76 & -4.18\tablefootmark{1} \\
PHL1811 & 47.5 & -44.8 & -358 & ... & 12.61 & 12.67 & ... & 13.78 & -3.80 \\
  &   &   & -261 & ... & 13.03 & 12.57 & 13.31 & 13.47 & -3.93 \\
  &   &   & -213 & 13.70 & 13.94 & 12.88 & 14.38 & ... & -3.30 \\
  &   &   & -169 & 12.90 & 13.12 & 13.43 & 13.76 & 14.10 & -3.30 \\
PKS2155-304 & 17.7 & -52.2 & -275 & ... & 12.22 & ... & ... & 13.56\tablefootmark{2}  & -4.15 \\
  &   &   & -138 & 12.73 & 13.18 & 12.62 & 13.78 & 13.48\tablefootmark{2} & -3.50 \\
Q2251+155 & 86.1 & -38.2 & -419 & 13.98 & 13.90 & 13.28 & 14.56 & ... & -2.98 \\
  &   &   & -372\tablefootmark{3} & 13.90\tablefootmark{3} & 14.53\tablefootmark{3} & 14.38\tablefootmark{3} & 14.78\tablefootmark{3} & 13.79\tablefootmark{3} & -3.68\tablefootmark{3} \\
  &   &   & -326 & 13.54 & 13.90 & 13.65 & 14.49 & 13.74 & -3.45 \\
QSO0045+3926 & 122.3 & -23.2 & -380 & 12.86 & 13.14 & ... & 14.15 & 13.12 & -3.36 \\
  &   &   & -327 & ... & 12.26 & ... & 13.46 & ... & -2.71 \\
  &   &   & -246 & 12.78 & 13.36 & 13.14 & 13.87 & 14.02 & -3.60 \\
  &   &   & -185 & 13.50 & 13.68 & 12.83 & 14.70 & 13.57 & -3.21 \\
QSO-B0026+129 & 114.6 & -49.2 & -187 & ... & 12.46 & ... & ... & ... & -3.30 \\
  &   &   & -289 & ... & 12.81 & 12.30 & 13.45 & ... & -3.03 \\
  &   &   & -241 & 12.78 & 12.73 & ... & 13.54 & ... & -2.85 \\
QSO-B2214+139 & 76.0 & -34.2 & -360 & ... & 13.22 & 12.65 & 13.42 & ...  & -3.18 \\
  &   &   & -327 & ... & 13.01 & ... & 13.26 & ...  & -3.95 \\
SDSSJ001224.01-102226.5 & 92.3 & -70.9 & -280 & 13.44 & 13.53 & 12.81 & 14.35 & ... & -3.15 \\
  &   &   & -243 & 13.22 & 13.53 & 13.48 & 13.96 & 14.17 & -3.37 \\
  &   &   & -180 & 13.40 & 13.27 & ... & 14.17 & 13.33 & -2.91 \\
SDSSJ004222.29-103743.8 & 115.1 & -73.4 & -147 & ... & 12.82 & ... & ... & ...  & -2.85 \\
  &   &   & -102 & 13.32 & 13.10 & ... & 14.09 & 13.40 & -2.83 \\
SDSSJ225738.20+134045.4 & 85.3 & -40.7 & -421 & ... & 13.00 & ... & 13.70 & ... & -3.17 \\
  &   &   & -374 & 13.72 & 13.48 & 13.37 & 14.25 & 13.79 & -2.86 \\
  &   &   & -329 & 13.33 & 13.48 & 13.25 & 14.17 & 13.80 & -3.25 \\
\pagebreak                 
RXJ1230 & 291.3 & 63.7 & 119 & 12.68 & 12.68 & ... & 13.88 & 12.83 & -3.07 \\
  &   &   & 299 & 13.06 & 12.77 & ... & 14.01 & 12.77 & -2.78 \\
SDSSJ121640 & 278.9 & 68.4 & 215 & 12.58 & 12.55 & ... & ... & ... & -3.02 \\
SDSSJ121850 & 276.9 & 71.5 & 110 & 13.14 & 13.25 & ... & 14.05 & ... & -3.15 \\
  &   &   & 151 & 12.94 & 12.93 & ... & 14.10 & ... & -3.07 \\
SDSSJ122018 & 281.6 & 68.3 & 208 & 13.00 & 12.85 & 13.10 & ... & ... & -2.93 \\
  &   &   & 273 & 12.87 & 12.72 & ... & ... & ... & -2.91 \\
SDSSJ122102 & 269.2 & 76.8 & 143 & 13.38 & 13.52 & ... & 14.15 & ...  & -3.20 \\
SDSSJ122312 & 280.5 & 71.5 & 172 & 13.10 & 12.85 & ... & 13.80 & ... & -2.82 \\
SDSSJ122317 & 281.1 & 71.1 & 188 & 12.80 & 12.58 & ... & 13.59 & ... & -2.84 \\
SDSSJ123304 & 293.1 & 62.0 & 125 & 12.71 & 12.59 & 12.68 & 13.35 & 13.30 & -2.96 \\
SDSSJ123426 & 290.6 & 69.9 & 199 & 13.00 & 12.95 & ... & ... & ... & -2.99 \\
  &   &   & 239 & 14.34 & 14.50 & ... & ... & ... & -3.25 \\
VV2006J131545 & 329.9 & 77.0 & 238 & ... & 12.93 & ... & ... & ...  & -2.68 \\

\hline
\tablefoottext{1}{log\,$n_{\rm H}$ from Si\,{\sc iv} and C\,{\sc iv}}\\
\tablefoottext{2}{from \citet{Collins04}}\\
\tablefoottext{3}{No reliable measurement\\
due to blending and low S/N}
\end{longtable}

\end{appendix}


\end{document}